\documentclass[10pt,conference]{IEEEtran}
\IEEEcompsocitemizethanks
\IEEEcompsocthanksitem

\newcommand{\Red}[1]{\textcolor[rgb]{1.00,0.00,0.00}{#1}}

\usepackage{amsmath,amsfonts}

\usepackage{balance}
\usepackage[]{collab}

\collabAuthor{yx}{red}{Yuxin Su}
\usepackage{cite}
\usepackage{amssymb}
\usepackage{graphicx}
\usepackage{textcomp}
\usepackage{xcolor}
\usepackage{subcaption, booktabs}
\usepackage{adforn}
\usepackage{listings}
\usepackage{color}
\usepackage{calligra}
\usepackage{multirow}
\usepackage[normalem]{ulem}
\usepackage{algorithm}  
\usepackage{algpseudocode}  

\usepackage{tcolorbox}
\usepackage{colortbl}

\usepackage[symbol]{footmisc}
\renewcommand{\thefootnote}{\fnsymbol{footnote}}

\def\BibTeX{{\rm B\kern-.05em{\sc i\kern-.025em b}\kern-.08em
    T\kern-.1667em\lower.7ex\hbox{E}\kern-.125emX}}
    
\makeatletter
\newcommand{\linebreakand}{
    \end{@IEEEauthorhalign}
    \hfill\mbox{}\par
    \mbox{}\hfill\begin{@IEEEauthorhalign}
}
\makeatother
    
\def\name{SemParser}
\usepackage[]{collab}
\collabAuthor{yt}{red}{YT'Note}

\definecolor{codegreen}{rgb}{0,0.6,0}
\definecolor{codegray}{rgb}{0.5,0.5,0.5}
\definecolor{codepurple}{rgb}{0.58,0,0.82}
\definecolor{backcolour}{rgb}{0.95,0.95,0.92}

\definecolor{viol}{RGB}{134,0,175}

\DeclareMathAlphabet{\mathcalligra}{T1}{calligra}{m}{n}
\usepackage{amssymb}

\newcommand\blfootnote[1]{%
  \begingroup
  \renewcommand\thefootnote{}\footnote{#1}%
  \addtocounter{footnote}{-1}%
  \endgroup
}

\begin{document}

\title{SemParser: A Semantic Parser for Log Analytics}

\author{\IEEEauthorblockN{Yintong Huo}
\IEEEauthorblockA{\textit{Computer Science \& Engineering Dept.} \\
\textit{The Chinese University of Hong Kong}\\
Hong Kong, China \\
ythuo@cse.cuhk.edu.hk}
\and
\IEEEauthorblockN{Yuxin\thanks{abc} Su\IEEEauthorrefmark{1}}
\IEEEauthorblockA{\textit{School of Software Engineering} \\
\textit{Sun Yat-sen University}\\
Zhuhai, China \\
suyx35@mail.sysu.edu.cn}
\linebreakand
\IEEEauthorblockN{Cheryl Lee}
\IEEEauthorblockA{\textit{Computer Science \& Engineering Dept.} \\
\textit{The Chinese University of Hong Kong}\\
Hong Kong, China \\
cheryllee@link.cuhk.edu.hk}
\and
\IEEEauthorblockN{Michael R. Lyu}
\IEEEauthorblockA{\textit{Computer Science \& Engineering Dept.} \\
\textit{The Chinese University of Hong Kong}\\
Hong Kong, China \\
lyu@cse.cuhk.edu.hk}
\\
}

\maketitle

\begin{abstract}
Logs, being run-time information automatically generated by software, record system events and activities with their timestamps. Before obtaining more insights into the run-time status of the software, a fundamental step of log analysis, called log parsing, is employed to extract structured templates and parameters from the semi-structured raw log messages.
However, current log parsers are all \textit{syntax-based} and regard each message as a character string, ignoring the semantic information included in parameters and templates.
\blfootnote{* Corresponding author.}

Thus, we propose the first \textit{semantic-based} parser \name~to unlock the critical bottleneck of mining semantics from log messages. It contains two steps, an end-to-end semantics miner and a joint parser. Specifically, the first step aims to identify explicit semantics inside a single log, and the second step is responsible for jointly inferring implicit semantics and computing structural outputs according to the contextual knowledge base of the logs. 
To analyze the effectiveness of our semantic parser, we first demonstrate that it can derive rich semantics from log messages collected from six widely-applied systems with an average F1 score of 0.985. Then, we conduct two representative downstream tasks, showing that current downstream models improve their performance with appropriately extracted semantics by 1.2\%-11.7\% and 8.65\% on two anomaly detection datasets and a failure identification dataset, respectively. We believe these findings provide insights into semantically understanding log messages for the log analysis community.


\end{abstract}

\section{Introduction}

The logging statements, which are put into the source code by developers, carry run-time information about software systems. By reading these logs, software system operators and administrators can monitor software status~\cite{chen2004failure}, detect anomalies~\cite{xu2009largescale, zhao2021empirical}, localize software bugs~\cite{chen2021pathidea}, or troubleshoot problems~\cite{amar2019mining} in the system.
The overwhelming logs, however impede developers from reading every line of log files as modern software systems get more complicated than before. Therefore, intelligent software engineering necessitates automated log analysis.

\begin{figure*}[tb]
    \centering
        {\includegraphics[width=\linewidth]{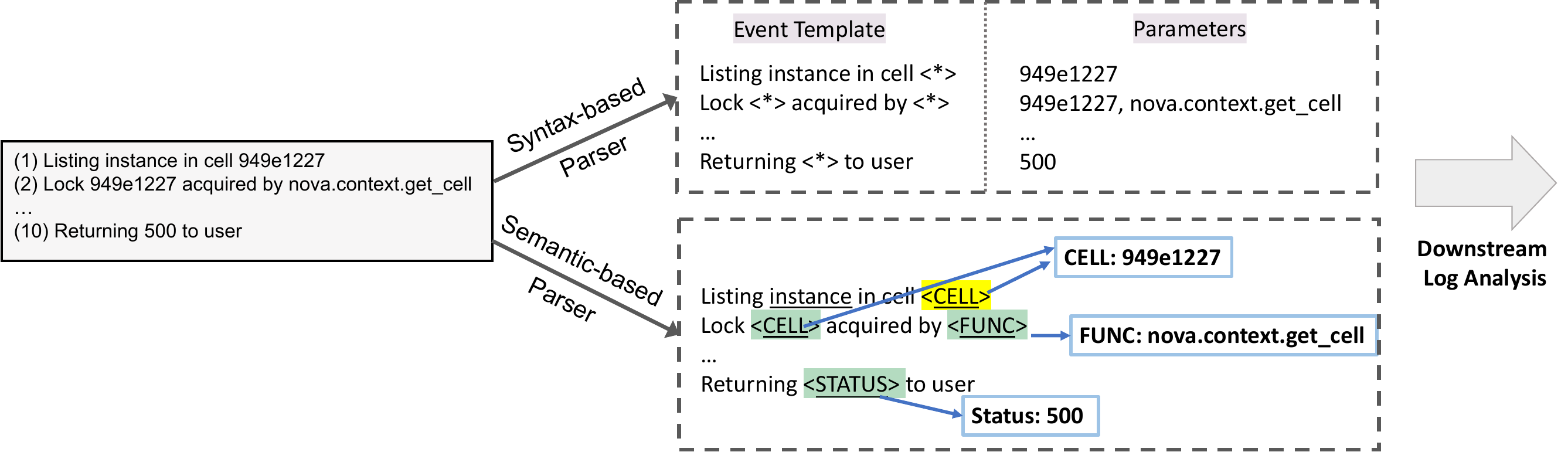}}
    \caption{Difference between syntax-based parsers and semantic-based SemParser. Logs are generated from OpenStack.}
    \label{fig:difference}

\end{figure*}

Basically, a log message is a type of semi-structured language comprising a natural language written by software developers and some auto-generated variables during software execution.
As most log analysis tools accept the structured input, the fundamental step for automated log analysis is log parsing. Given a raw message, a log parser recognizes a set of fields (e.g., verbosity levels, date, time) and message content, while the latter being represented as structured event templates (i.e., constants) with corresponding parameters (i.e., variables). 
For example, in Figure~\ref{fig:difference} (up), ``Listing instance in cell $<$*$>$'' is the template describing the system event, and ``949e1227'' corresponds to the parameter indicator ``$<$*$>$'' in the template.

Although automatic log parsing is full of challenges, researchers have made progress leveraging statistical and history-based methods.
For instance, SLCT~\cite{vaarandi2003data} and LFA~\cite{nagappan2010abstracting} constructed log templates by counting the number of historical frequently-appearing words while Logram~\cite{dai2020logram} considered frequent n-gram patterns. LogSig~\cite{tang2011logsig} and SHISO~\cite{mizutani2013incremental} encoded the log by word pairs and words length, respectively, then applied the clustering algorithm for partitioning.~\cite{chu2021prefix} adopted the idea of probabilistic graph for parsing. The most widely-used parser in industry, Drain~\cite{he2017drain}, formed log templates by traversing leaf nodes in a tree. However, we argue that all current parsers are \textit{syntax-based} with superficial features (e.g., word length, log length, frequency), and they have limited high-level semantic acquisition.
In this paper, we classify the limitations into a three-level hierarchy.



The first is paying inadequate attention to individual \textit{informative tokens}. 
Taking the first log in Figure~\ref{fig:difference} as an example, the parameter (i.e., 949e1227) and technical concepts (i.e., instance, cell) are noteworthy, comparing with other preposition words (e.g., in). Syntax-based log parsers only distinguish parameters and templates but treat each log message as a sequence of characters without paying attention to special technical concepts.
A previous study~\cite{li2018studying} found that technical terms and topics in logs are informative by studying six large software systems. Therefore, both the parameters and domain terms should be localized for log comprehension.

Secondly, the \textit{semantics within a message} should be noticed. While humans seldomly use digits or character strings (e.g., 949e1227) in communication, parameters in the log message are important with specific meaning. Unfortunately, syntax-based parsers regard each parameter as a meaningless character string.
Intuitively, a parameter in a log is used to specify another technical concept in the log. For example, from the first log in Figure~\ref{fig:difference}, we understand that the token ``949e1227'' refers to another token ``cell'', so ``949e1227'' is a cell ID. In this way, exploiting such intra-message semantics benefits the understanding of parameters. 



Thirdly, the \textit{semantics between messages} are missing. 
All previous parsers process each log message independently, ignoring the inter-message relation between logs. However, historical logs can provide domain knowledge of a parameter, helping resolve the implicit semantics of the same parameter in subsequent logs. In Figure~\ref{fig:difference}, though the second log does not explicitly disclose the semantics of parameter ``949e1227'', we know it refers to a cell based on the historical information provided in the first log. As parameters rarely appear in daily language, mining semantics from log messages is distinct from understanding common language.

Some studies notice the above limitations and attempt to mitigate them. 
LogRobust~\cite{zhang2019robust} assigned weights towards each token based on the TF-IDF value when encoding logs to reveal informative tokens. 
This approach tends to assign the rare word with a high attention weight, but common technical terms can also be illuminating.
For semantic mining, Drain~\cite{he2017drain}, LKE~\cite{fu2009execution}, MoLFI~\cite{messaoudi2018search} and SHISO~\cite{mizutani2013incremental} used regular expressions to recognize block ID, IP address, and number when parsing HDFS datasets. However, designing human handcrafted rules requires tedious effort and suffers from system migrations. It is impossible to exhaust all possibilities, so the rules can only cover a fairly limited part of the logs.
Besides, these regular expressions cannot distinguish polysemy of parameters. For instance, the variable ``200'' refers to the return code if the system makes REST API calls, but it may also represent a thread identifier (TID) in Spark. 
Moreover, although text mining approaches~\cite{blei2003latent, he2017unsupervised} try to mine semantics from human language, they cannot understand the variables with specific meaning in log messages. 
As shown in the last log in Figure~\ref{fig:difference}, the serious information omissions and misunderstanding of the erroneous status code ``500'' will accumulate as the scale of the parsed logs increases, and ultimately hinder the further anomaly detection task, making it difficult to accomplish the goal of avoiding incidents and ensuring system reliability.


To tackle the aforementioned complicated but critical limitations, we propose a novel \textit{semantic-based} log parser,~\textbf{\name}, the first work to target parsing logs with respect to their semantic meaning. 
We first define two-level granularities of semantics in logs, \textit{message-level} and \textit{instance-level semantics}. Message-level semantics refers to identifying technical concepts (e.g., cell) within log messages (underscored in Figure~\ref{fig:difference}), while instance-level semantics means resolving what the instance (i.e., parameters) describes.
Then, we design an end-to-end semantics miner and a joint parser that can not only recognize the templates of given logs, but also extract explicit semantics inside a log and the implicit inter-log semantics.
Specifically, the end-to-end semantics miner is devised to recognize the semantics of messages (e.g., concepts like ``instance'' and ``cell''), and explicit semantics of instances (e.g., ``949e1227'' refers to ``cell''). In this way, the noteworthy tokens and explicit semantics of parameters are obtained to break the first and second limits, respectively.
The joint parser then infers the implicit semantics of parameters with the assistance of domain knowledge acquired from prior logs, mitigating the third limitation of missing inter-log relation.
Figure~\ref{fig:difference} illustrates the major difference between the syntax-based parsers and the proposed \name, where the explicit semantics is highlighted in yellow and implicit semantics is highlighted in green.
Obviously, not only can~\name~play the role of an accurate log-template extractor as syntax-based parsers, but also it can provide additional and structured semantics to promote downstream analysis.

We conduct an extensive study to investigate the performance of~\name~on six system logs from two perspectives: (1) the effectiveness for semantic mining; (2) its effectiveness on two typical log analysis downstream tasks. 
The experimental results demonstrate that our approach can capture semantics more accurately, which achieves an average F1 score of 0.985 in semantic mining, and that it outperforms state-of-the-art log parsers by the average of 1.2\% and 11.7\% on two anomaly detection datasets and 8.65\% on a failure identification dataset. These powerful results reveal the superiority of \name~and emphasize the importance of semantics in log analytics, especially when the software systems we handle are more complicated than ever before.


In summary, the contribution of this paper is threefold:
\begin{itemize}
    \item To our best knowledge, \name~is the first semantic-based parser capable of actively capturing message-level and instance-level semantics from logs, as well as actively collecting and leveraging domain knowledge for parsing.
    \item We evaluate \name~with respect to its semantic mining accuracy on six system logs, demonstrating our framework could effectively mine semantics from logs.
    \item We also employ \name~on the failure identification and anomaly detection tasks, and the promising results reveal the importance of semantics in the log analytics field.
\end{itemize}


\section{Problem Statement}\label{sec:statement}


This paper focuses on parsing logs with respect to semantics, which could further be decoupled into message-level semantics and instance-level semantics. 
Message-level semantics are defined as a set of \textit{concepts} (i.e., technical terms) appearing in log messages, such as ``cell''. 
We use the term $instance$ \footnote{The term ``instance'' is rather closed to the ``parameters'' or ``variables'' in the syntax-based parser. One concept can be instantiated by multiple instances.} to denote variables in log messages, then the instance-level semantics are represented by a set of \textit{Concept-Instance pairs (CI pairs)}, which describe the concept that the instance refers to, such as (cell, 949e1227). A \textit{Domain Knowledge database} maintains a list of detected CI pairs from historical logs.
After obtaining \textit{instances}, \textit{concepts} and \textit{CI pairs} from a log message, we replace the instances with their corresponding concepts and name the new message as \textit{conceptualized template}. 

The semantic parser task can be regarded as following. Given a log message\footnote{The log message refers to log content without fields in this paper by default.}, the structural output is composed of a conceptualized template $T$, a set of $CI$ pairs $CI=\{(c_0,i_0), ..., (c_n,i_n)\}$, as well as other orphan concepts $OC=\{oc_0,...,oc_j\}$ and orphan instances $OI=\{oi_0,...oi_k\}$ which cannot be paired with each other.
\begin{figure}[tb]
    \centering
        {\includegraphics[width=\linewidth]{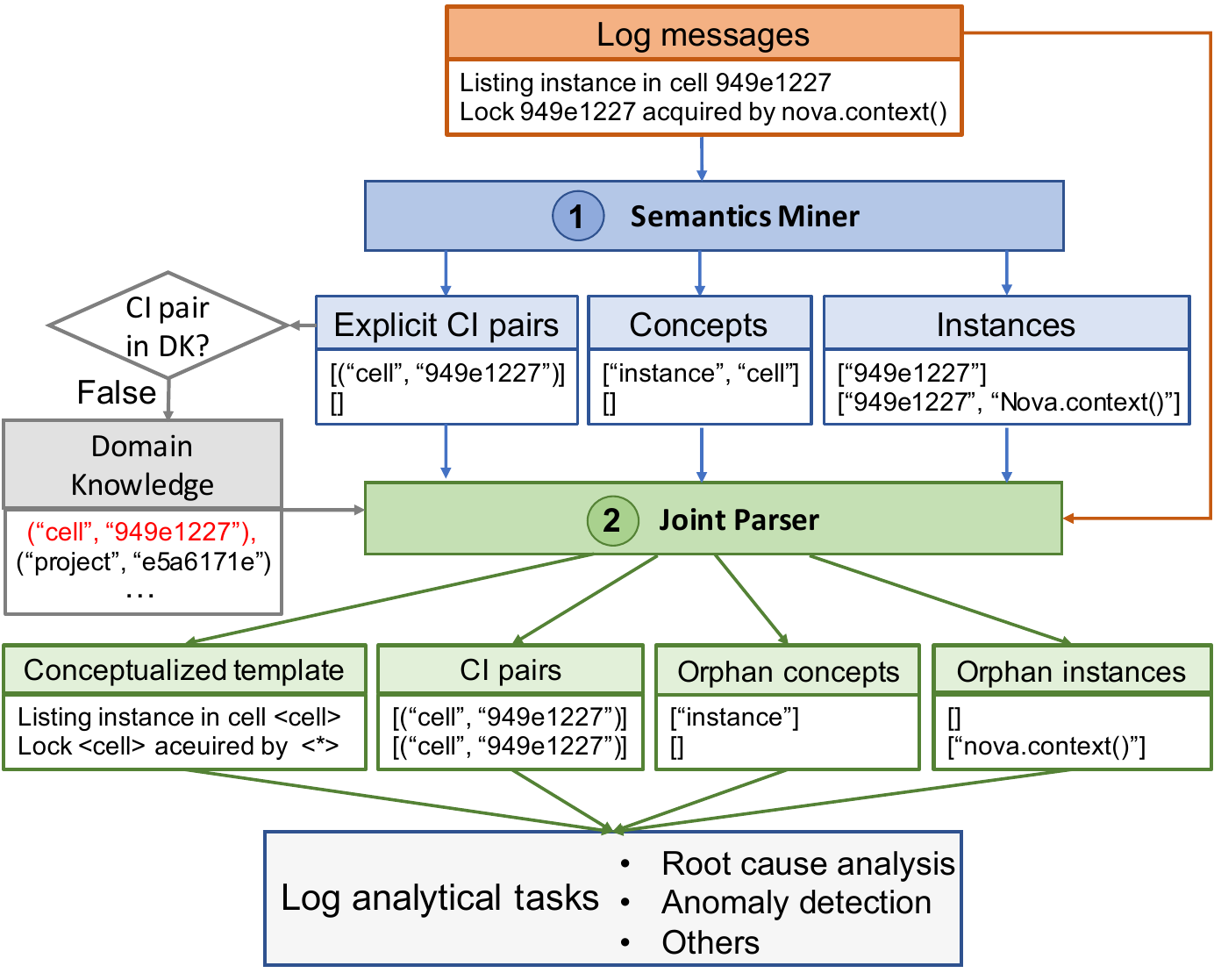}}
    \caption{The pipeline of~\name.}
    \label{fig:pipeline}

\end{figure}

\begin{figure*}[tb]
    \centering
        {\includegraphics[width=0.95\linewidth]{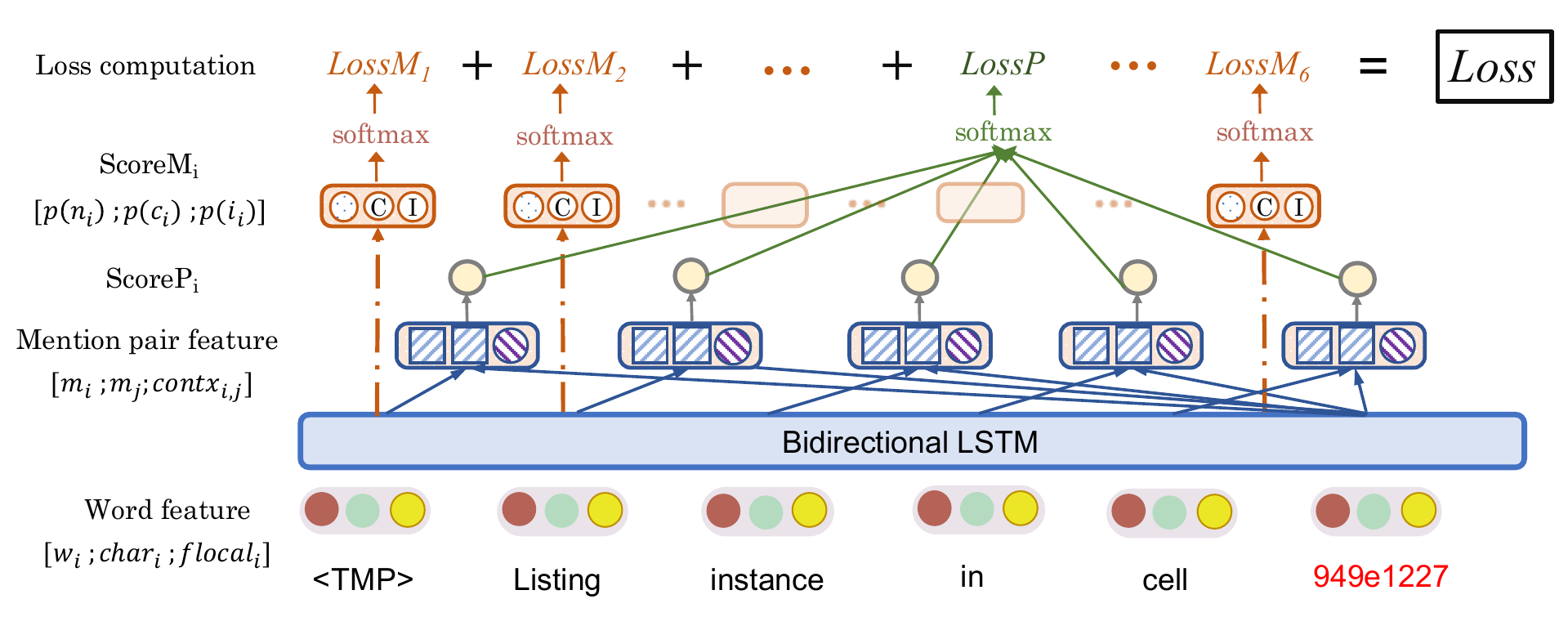}}
    \caption{The architecture of semantics miner.}
    \label{fig:framework}

\end{figure*}

\section{Methodology}\label{sec:overview}
\subsection{Overview of \name}\label{subsec: overview}

Our framework is composed of two parts, an end-to-end semantics miner and a joint parser. 
In Figure~\ref{fig:pipeline}, we use an example to illustrate how our framework processes log messages.
To begin with, log messages are sent to the semantic miner for acquiring template-level semantics (i.e., \textit{concepts}) and explicit instance-level semantics (i.e., \textit{explicit CI pairs}) of each log independently. This step particularly solves the first two stated challenges.
The unseen explicit CI pairs will be added to the \textit{Domain Knowledge database} to keep the knowledge updated. Moreover, to uncover potential implicit semantics from domain knowledge, \textit{instances} in log messages are kept. Hence, the challenge of missing inter-log relations are addressed.

Following that, the joint parser receives outputs from the semantics miner, taking charge of implicit semantics inference with the help of domain knowledge. The newfound implicit instance semantics, coupled with the explicit one, form the instance-level semantics, denoted as \textit{CI pairs}. The remaining concepts and instances which cannot be paired with each other are stored as \textit{orphan concepts} and \textit{orphan instances}, respectively. Besides, the \textit{conceptualized templates} are derived by replacing instances with their corresponding concepts (if available), or ``$<$*$>$'' for else. The final structural outcome of~\name~consists of \textit{conceptualized templates}, \textit{CI pairs}, \textit{orphan concepts}, as well as \textit{orphan instances}.

As the first and fundamental step for log analysis,~\name~could facilitate general downstream log analysis tasks. We will introduce details of the semantics miner and the joint parser in the following two subsections. Then, two typical downstream applications will be displayed in Section~\ref{sec:results}.



\subsection{End-to-end semantics miner}\label{subsection:sminer}
Semantics miner aims to mine semantics on both the instance-level and the message-level. To acquire a set of explicit \textit{concepts},  \textit{instances} , and \textit{CI pairs} within a log message, we model the task into two sub-problems: finding CI pairs and classifying each token into a type in $\{concept, instance, none\}$. 
As shown in Figure~\ref{fig:framework}, an end-to-end model with three modules is proposed to solve the two sub-tasks simultaneously. 
First, a log message is fed into a \textit{Contextual Encoder} for acquiring context-based word representation.
Then, the contextualized words are separately used in \textit{Pair Matcher} and a \textit{Word Scorer} for extracting CI pairs and determining the type of each word, respectively. As the total loss is the sum of the Pair Matcher loss and Word Scorer loss, the model is forced to learn from both sub-tasks jointly. We elaborate on the details of the three modules as below.




\subsubsection{Contextual encoder}

Intuitively, log messages can be regarded as a special type of natural language due to its semi-structured essence of mixing unstructured natural language and structured variables.

Motivated by the success of long short-term memory networks (LSTM) across natural language processing tasks~\cite{sundermeyer2012lstm} (e.g., machine translation, language modeling), we design an bi-directional LSTM-based network (bi-LSTM)~\cite{graves2013speech} to capture interactions and dependencies between words in log messages. 

However, it is not practical to directly feed the word embeddings into the LSTM network because of the severe out-of-vocabulary (OOV) problem, which is due to the large portion of customized words in log messages (e.g., function names, cell ID, request ID), resulting drastic performance degradation. 
To solve the problem, we devise two additional features associated with word representations.
Firstly, inspired by previous findings that character-level representation helps exploit sub-word-level information~\cite{ma-hovy-2016-end}, we adapt a Convolution Neural Network (CNN) to extract character-level features of each word. 
Secondly, following several studies~\cite{huang2015bidirectional, chiu2016named} that leveraged local features for sequential representations, we also deliberate a set of local features for each word concerning its shape, length, and other morphological features.

The word representations $word_i$, character representations $char_i$, as well as the local features $f_{i}^{local}$, are concatenated as word features and fed into the bi-LSTM indicated in Equation~\ref{Equ:lstm}. Afterward, the hidden state of bi-LSTM is used as the contextual embedding for each word.
\begin{eqnarray}
m_i = LSTM([word_i;char_i;f_{i}^{local}])
\label{Equ:lstm}
\end{eqnarray}


\subsubsection{Pair matcher}
This module is designed for acquiring explicit instance-level semantics.
Numerous studies focus on identifying key elements in texts and classify them into several categories by assigning each word into one of the pre-defined categories. For example, the combination of bi-LSTM and Conditional Random Field (CRF) is deployed to identify 100 log entities (e.g., IP address, identifier) in log messages~\cite{shetty2021neural}, or uncover 20 software entities (e.g., class name, website) in software forum discussions~\cite{tabassum-etal-2020-code}.
However, such token classification-based framework relies on a \textit{closed-world assumption} that all categories are known in advance. The assumption makes sense when dealing with a specific and small system with limited concepts. Unfortunately, it will break down if we want to migrate the approach across software systems, or the system we are facing is huge and sophisticated.

To get over the closed-world assumption limitation, the pair matcher is required to discern the (\textit{concept}, \textit{instance}) pairs between words in a log message. We abstract this problem as a multi-classifier problem: for each word $w_i$ in a sentence $S = w_1, w_2, ..., w_n$, the matcher determines what previous word $w_j (0 \leq j < i)$ does the word $w_i$ refer to\footnote{We add a dummy word $<$TMP$>$ ($w_0$) to indicate the word does not refer to any of the previous word in the message (e.g., in). }. 


To achieve the goal, we rank the confidence score of each word pair candidate $(w_i,w_j), \forall 0\leq j < i$, which is determined by a feed forward neural network $FFNN_a$ as in Equation~\ref{Equ:score}. 
Intuitively, if a word ``is'' exists between $w_i$ and $w_j$, the pair has a higher probability formed by the two words, so we consider the interval context between the candidate pair $(w_i,w_j)$ as the average word embedding value between the pair, denoted as $contx_{i,j}$. In summary, we construct the pair-level features $f_{i,j}^{pair}$ for scoring by concatenating contextual representation (i.e., $m_i,m_j$) obtained from last step, as well as the abovementioned interval context $contx_{i,j}$, as shown in Equation~\ref{Equ:pair-feature}. 
\begin{eqnarray}
ScoreP_i(i,j)& =& FFNN_a(f_{i,j}^{pair}) \label{Equ:score}\\
f_{i,j}^{pair}&=&[m_i; m_j; contx_{i,j}] \label{Equ:pair-feature}
\end{eqnarray}

Figure~\ref{fig:framework} shows a simple case for matching pair for the red word $w_5$. After acquiring contextual word representations from the contextual encoder, we form the pair-level feature for each word pair in $\{(w_5,w_4), (w_5,w_3)...(w_5,w_0)\}$. These pair features will be scoring by a softmax function on top of a feed forward neural network for loss computation.




\subsubsection{Word scorer}
Apart from the pair matcher, we also design a word scorer to determine whether each token is a \textit{concept}, \textit{instance} or \textit{neither of both}. The token's category is crucial for two reasons. First, the message-level semantics can be perceived via extracted concepts.
Second, we notice that some instance-level semantics cannot be resolved via the pair matcher if the instance's corresponding concept does not occur in a single message (e.g., the second log in Figure~\ref{fig:difference}), which we call \textit{implicit instance-level semantics}. In this case, we need to store the \textit{instances} for further processing. To this end, we devise the word scorer with a feed-forward neural network $FFNN_b$ to learn the possibility of three types for each token. The score is computed as follows:
\begin{eqnarray}
ScoreM_{i} = FFNN_b(m_i)
\label{Equ:mention-score}
\end{eqnarray}
Afterwards, the possibility of three categories will pass through a softmax layer for normalization before computing loss.

\subsubsection{Loss function}
Multi-task learning (MTL) is a training paradigm that trains a collection of neural network models for multiple tasks simultaneously, leveraging the shared data representation for learning common knowledge\cite{caruana1997multitask, shetty2021neural}. The fruitful achievements of MTL motivate us to train pair matcher and word scorer simultaneously by aggregating their losses.
Therefore, the total cost of semantics miner is defined as:
\begin{eqnarray}
cost= \sum_i CELoss({P_i}') + \sum_i CELoss({M_i}')
\end{eqnarray}
where ${P_i}'$ and ${M_i}'$ denotes the outputs of $ScoreP_i$ and $ScoreM_i$ after passing a softmax layer, respectively. Here, we adopt Cross Entropy Loss (i.e., CELoss) as the loss function due to its numerical stability.
By minimizing the cost, the model naturally learns the pairs and the word types for each token with shared contextual representations generated from bi-LSTM network.

In the inference, for each word, we regard the highest probability of its pairs and its type score as the final results.

%
\subsection{Joint parser}\label{subsection:jparser}
The joint parser leverages concepts, instances, and CI pairs obtained from the end-to-end semantics miner, as well as log messages to deal with                                                                                                                                                                                                                    : (1) uncovering implicit instance-level semantics using domain knowledge; and (2) semantic parsing log messages. The next sections go into the specifics.

\subsubsection{Implicit instance-level semantics discovery} 
We apply a novel domain knowledge-assisted approach to resolve the implicit instance-level challenge of concepts and instances not coexisting in one log message. Naturally, suppose we have recognized a CI pair in historical logs, then we are able to identify the semantics of such instance in the following logs, even though the following logs do not explicitly contain such pair information. 

The knowledge-assisted approach maintains a high-quality domain knowledge database when processing logs by incorporating newly discovered CI pairs acquired from the semantics miner. To guarantee the quality of the domain knowledge, we only add the superior CI pairs, which are defined by \textit{if and only if there is a concept and an instance in the predicted pair}.
The joint parser examines whether the orphan instances have their corresponding concepts in the high-quality knowledge base, to uncover implicit CI pairs. As a result, fresh CI pairs of the log messages are stored if found. In such a way, we merge the explicit CI pairs and new implicit CI pairs into the final CI pairs. Details are in Algorithm~\ref{algo:implicit-miner}.

\begin{algorithm}[h]  
\small
  \caption{Implicit instance-level semantics discovery}  
  \label{algo:implicit-miner}
  \begin{algorithmic}[1]  
    \Require Log message $M=m_0,...,m_n$, instance indices $I=[i_0,...i_j]$, concept indices $C=[c_0,...c_k]$, explicit CI pair indices $P=[(s_0,t_0),...,(s_u,t_u)]$
  \Ensure Instances $I'$, Concepts $C'$, CI pairs $P'$
\State $P' = []$
\State $C' = []$
    \ForAll {$p$ such that $p\in P$}  
    \If{$p$ contains 1 instance $cur_I$ and 1 concept $cur_C$}
      \State{DomainKnowledge.add($M[cur_C]$,$M[cur_I]$)}
      \State{$I$.\textsc{remove}($cur_I$)}
      \State{$C$.\textsc{remove}($cur_C$)}
      \EndIf
    \EndFor;  
    
    \ForAll{$i$ such that $i \in I$}
    \If{\textsc{FindConceptFromDomainKnowledge}($M[i]$)}
    \State{$P'$\textsc{.append}([newfound concept, $M[i]$])}
    \State{$C'$.\textsc{append}(newfound concept)}
    \State{$I$.\textsc{remove}($i$)}
    \EndIf
    \EndFor
\State $I'$ = \textsc{IndexToWord}$(I)$
\State $C'$ += \textsc{IndexToWord}($C$);
\State $P'$ += \textsc{IndexToWord}$(P)$

  \end{algorithmic}  
\end{algorithm}  

\subsubsection{Semantic parsing}

As a semantic parser,~\name~is able to extract the template for a given log message obeying two rules:
\begin{itemize}
    \item For the instance in CI pairs, replacing the instance with the token $<$concept$>$ of its corresponding concept.
    \item For the orphan instances, replacing the instance with a dummy token $<$*$>$ as syntax-based parsers do.
\end{itemize}

The rules are straightforward but reasonable.
Compared to other technical terms or common words, instances (e.g., ID, number, status) are more likely to be variables in logging statements automatically generated by software systems. As the retrieved template takes in concepts, we name it ``conceptualized template'' instead of the vanilla template with only $<$*$>$ representing parameters.

Finally, the conceptualized template, CI pairs, orphan concepts, as well as orphan instances are the structured outputs of our~\name. The results are extensible for a collection of downstream tasks, and we will elaborate them later.

\section{\name~Implementation}\label{sec:details}

\subsection{Dataset annotation}
We implement the~\name~framework on a public dataset~\cite{cotroneo2019bad} containing log messages collected from OpenStack for training.
Considering that it is labor-intensive to annotate a large dataset in a real-world scenario, we randomly sample 200 logs from the dataset for human annotation, with the sample rate of 0.05\%. A practical model should be able to learn from a small amount of data. The trained model from such data is named the ``base model'' for further evaluation.

All annotation is carried out as follows.
For each log, we invite two post-graduate students experienced in OpenStack to independently manually label: (1) whether a word is a concept, instance, or neither of both; and (2) the explicit CI pairs within a sentence.
If the two students provide the same answer for one log, the answer will be regarded as the ground-truth for training; otherwise another student will join them to discuss until a consensus is reached. The inter-annotator agreement~\cite{cohen1960coefficient} before adjudication is 0.881.
Finally, we remove the sentences without any CI pair annotation to mitigate the sparse data problem, yielding 177 labeled messages for training the semantics miner.

\subsection{Pre-trained word embeddings}

Although existing pre-trained word embeddings show the large success in representing semantics of words, it is not appropriate for understanding logs. Log message is a very domain-specific language, where the words have quite distinct semantics from daily life.
Hence, we train domain-specific word embeddings on a representative cloud management system, OpenStack corpus. The corpus is made up of 203,838 sentences crawled from its official website. We train the pervasive skip-gram model~\cite{mikolov2013distributed} on Gensim~\cite{rehureklrec} for ten epochs and set the word embedding dimension to be 100.

\subsection{Implementation details}
When implementing the model, we set the character-level embedding dimension to be 30. 
We select the two-layer deep bi-LSTM with a hidden size of 128. The model is trained for 30 epochs\footnote{The model converges within 30 epochs.} with an initial learning rate of 0.01. The learning rate decays at the rate of 0.005 after each epoch.
It takes one hour for training, and the trained model occupies only 25 MB. \name~runs 25 messages per second in a single batch and single thread during inference.
\section{Evaluation}\label{sec:results}
We evaluate~\name~from two perspectives, the ability of semantic mining and the usefulness in downstream tasks, with three research questions:
\begin{itemize}
    \item RQ1: How effective is the~\name~in mining semantics from logs?
    \item RQ2: How effective is the~\name~in anomaly detection?
    \item RQ3: How effective is the~\name~in failure identification?
\end{itemize}

\subsection{Experiment details}\label{subsec:dataset}

\subsubsection{RQ1--Semantic mining}\label{subsec:dataset-sminer}

\textbf{Dataset.} LogHub~\cite{he2020loghub} is a repository of system log files for research purposes, which has been used by plenty of log-related studies~\cite{liu2019logzip, chen2021experience, lin2016log}. We manually label six representative log files for semantic mining evaluation ranging from distributed, operating, and mobile systems.
The dataset has a total of six different system log files with 12,000 log messages and 20,636 annotated CI pairs. Details are shown in Table~\ref{tab:loghub-stat}, where \# Logs, \# Pairs, \# Temp., and Unseen denotes the number of log messages, CI pairs, log templates, and the percentage of \textit{unseen templates} in the test set, respectively.

\textbf{Settings.} 
As~\name~is an semantic-based parser, we consider its semantic mining ability for evaluating how effective is it when mining instance-level semantics from log messages. Specifically, given a log message, we report the correct proportion of the model's extracted CI pairs (Precision), the proportion of actually correct positives extracted by the model (Recall), and their harmonic mean (F1 score). 
As we hope the model could learn semantics from small samples, we fine-tune the base model (i.e., train from Section~\ref{sec:details}) on a small dataset 50 randomly sampled logs for each system and evaluate the performance on the remaining 1,950 logs. 

\begin{table}
\footnotesize
\centering
\caption{Statistics of dataset for semantic mining.}
\begin{tabular}{c|c|c|c|c|c}
\toprule
\textbf{System type} & \textbf{System}    & \textbf{\#Logs} & \textbf{\#Pairs} & \textbf{\#Temp.} & \textbf{Unseen} \\ \toprule
Mobile system & Android   & 2,000 & 6,478 & 166  & 82.8\% \\ \midrule
Operating system & Linux     & 2,000 & 2,905 & 118  & 86.8\% \\ \midrule
\multirow{5}{*}{Distributed system} & Hadoop    & 2,000 & 2,592 & 14 & 84.6\%  \\ 
  & HDFS      & 2,000 & 3,105 & 30  & 	47.0\% \\ 
  & OpenStack & 2,000 & 4,367 & 43  & 52.3\% \\ 
  & Zookeeper & 2,000 & 1,189 & 50  & 75.9\% \\ \bottomrule
\end{tabular}
\label{tab:loghub-stat}
\end{table}

\subsubsection{RQ2--Anomaly detection}\label{subsec:dataset-ad}



\textbf{Dataset.} We evaluate the anomaly detection performance on two datasets. 
(1) We first follow the previous studies to evaluate in the HDFS~\cite{xu2009detecting} dataset, which includes log messages by running map-reduce tasks on more than 200 nodes.
(2) The second F-Dataset~\cite{cotroneo2019bad} is initially created for investigating software failures by injecting 396 failure tests in major subsystems of the widely used cloud computing platform OpenStack, covering 70\% of bug reports in the issue tracker.
For each failure injection test, the authors all \textit{log data} in major subsystems, the \textit{labeled anomaly log messages}, as well as the exception raised by a service API call named as \textit{API Error}, such as ``server create error''. 
Statistics of both datasets are shown in Table~\ref{tab:stat-ad}.

\begin{table}[t]
\small
    \centering
        \caption{Statistics of anomaly detection datasets.}
    \begin{tabular}{l||c|c}
    \toprule
    Dataset & \#Message  & Anomaly rate\\
    \midrule
    HDFS dataset & 11,175,629 & 3\% \\
    F-Dataset & 1,318,860 & 0.22\% \\
         \bottomrule
    \end{tabular}
    \label{tab:stat-ad}

\end{table}

\textbf{Settings.} In the anomaly detection task, the detector predicts whether anomalies exist within a short period of log messages (i.e., session). 
Motivated by previous studies~\cite{he2016experience,chen2021experience}, we decouple the anomaly detection framework into two components, a \textit{log parser} to generate templates, and a \textit{detection model} to analyze template sequences in a session.
A dependable parser should perform well as a foundational processor for log analysis, regardless of the down-streaming detection model used. In our experiments, we compare the performance of different baseline parsers under various anomaly detection techniques.

Specifically, we compare~\name~to the following log parsers as baselines: (1) \textbf{LenMa~\cite{shima2016length}.} This online parser encodes each log message into a vector, where each entry refers to the length of the token. Then, it parses logs by comparing the encoded vectors; (2) \textbf{AEL~\cite{jiang2008abstracting}.} This paper devises a set of heuristic rules to abstract values, such as ``value'' in ``word=value''; (3) \textbf{IPLoM~\cite{makanju2009clustering}.} IPLoM partitions event logs into event groups in three steps: partition by the length of the log; partition by token position; and partition by searching for bijection between the set of unique tokens; (4) \textbf{Drain~\cite{he2017drain}.} It leverages a fixed depth parse tree with heuristic rules to maintain log groups. Its ability to parse logs in a streaming and timely manner makes it popular in both academia and industry.

We also reproduce four widely-applied anomaly detection models as following: (1) \textbf{DeepLog}~\cite{du2017deeplog} employed a deep neural network, LSTM, to conduct anomaly detection and fault localization on logs, taking the context information into account; (2) To handle the ever-changing log events and sequences during the software evolution, \textbf{LogRobust}~\cite{zhang2019robust} detected anomaly detection by an attention-based bi-LSTM network. The attention mechanism allows the model to learn the different importance of log events; (3) \textbf{CNN}~\cite{lu2018detecting}  is also utilized to detect anomalies in big data system logs inspired by its benefits in general NLP analysis; and
(4)\textbf{Transformer.}~\cite{nedelkoski2020self} detected anomalies in logs via the Transformer encoder~\cite{vaswani2017attention} with a multi-head self-attention mechanism, allowing the model to learn context information.

When conducting experiments, we feed parsing results from log messages into different models. Different from previous work~\cite{du2017deeplog,zhang2019robust,lu2018detecting,nedelkoski2020self} that only employs templates to form the input sequence $x_0, x_1, ...,x_m$ where $x_i$ refers to the $i^{th}$ message in the sequence, we equip the sequence with extracted semantics. Specifically, for each log message in the sequence, we concatenate template, concepts, instances as follows:
\begin{eqnarray}
\tilde{x}=[template; <SEP>; sem_0; sem_1;...;sem_n]\\
sem_i=[concept_i; instance_i].
\end{eqnarray}
To specify the corresponding relationship within a CI pair, we concatenate the concept and instance in $sem_i$. Otherwise, an $<$NIL$>$ token replaces another half pair, indicating the orphan situation. A special $<$SEP$>$ token is used to separate template and semantics.
Afterwards, the sequence $\tilde{x_0}, \tilde{x_1}, ...,\tilde{x_m}$ containing $m$ messages will be fed into the model for prediction.
Following previous anomaly detection work~\cite{du2017deeplog, zhang2019robust, lu2018detecting, nedelkoski2020self}, we use Precision, Recall, and F1 as the evaluation metrics.

\subsubsection{RQ3--Failure identification}\label{subsec:dataset-rca}

\textbf{Dataset.} While anomaly detection identifies present faults from logs, failure identification looks deeper into the problems and identify what type of failure occurs. 
To make the F-Dataset appropriate for failure identification, we utilize the labeled anomaly log messages and their corresponding API error in each injection test as the input and ground-truth. Entirely, we collect 405 failures with 16 different types of API errors. With the splitting training ratio of 0.5, we obtain 194 and 211 failures for the train and test set, respectively. Typical API errors include ``server add volume error'', ``network delete error'' and so on.

\textbf{Settings.}
In this paper, we formulate the failure identification task as follows: given the \textit{anomaly log messages} from one injection test in F-Dataset, the model is required to determine what \textit{API error} emerges. Similar to the anomaly detection task, we also compare the performance of different baseline parsers associated with several log analysis models (i.e., DeepLog, LogRobust, CNN, and Transformer).
The only difference is that we change the node number of the last prediction layer of the above-mentioned techniques from 2 to 16 to make it a 16-class classification task for 16 error types in the dataset.

Recall@k is widely used in recommendation systems to assess whether the predicted results are relevant to the user(s)~\cite{covington2016deep, osadchiy2019recommender}. Similarly, we are also interested in whether top-k recommended results contain the correct API error. Hence, we report the Recall@k rate as the evaluation metric.

\begin{table}[t]
\footnotesize
    \centering
        \caption{Sample log messages and ground-truth templates.}

    \begin{tabular}{c|c}
    \toprule
       Log  &  After Scheduling: PendingReds:1 CompletedReds:0 ...\\
       GT-Template  & After Scheduling: PendingReds:$<$*$>$ CompletedReds:\Red{0} ... \\
    \midrule
        Log & TaskAttempt: [attempt\_14451444] using containerId ... \\
        GT-Template & TaskAttempt: [\Red{attempt\_$<$*$>$}] using containerId ... \\
    \bottomrule
    \end{tabular}
    \label{tab:discussion-gt}

\end{table}
 
\begin{table*}[t]
\footnotesize
    \centering
        \caption{Experimental results of mining semantics from logs.}
    \begin{tabular}{l||c|c|c|c|c|c}
    \toprule
    & \multicolumn{6}{c}{System}\\
        \cmidrule{2-7}
     & \multicolumn{1}{c}{Andriod} & \multicolumn{1}{c}{Hadoop} & \multicolumn{1}{c}{HDFS} & \multicolumn{1}{c}{Linux} & \multicolumn{1}{c}{OpenStack}  & \multicolumn{1}{c}{Zookeeper}\\
    Framework & P\quad |\quad R\quad |\quad F1 & P\quad |\quad R\quad |\quad F1 & P\quad |\quad R\quad |\quad F1 & P\quad |\quad R\quad |\quad F1 & P\quad |\quad R\quad |\quad F1 & P\quad |\quad R\quad |\quad F1 \\
    \midrule
    \name & 0.951 0.935 \textbf{0.943} & 0.993 0.978 \textbf{0.985} & 1.000 1.000 \textbf{1.000} & 0.998 0.977 \textbf{0.987} & 0.999 0.998 \textbf{0.999} &  1.000 0.989 \textbf{0.995}\\
    - w/o $F_{char}$ & 0.981 0.909 \textbf{0.943} & 0.988 0.953 0.970 & 1.000 0.998 0.999 & 0.995 0.957 0.976 & 0.995 0.989 0.992 &  0.993 0.987 0.990\\
    - w/o $F_{local}$ & 0.979 0.858 0.915 & 0.993 0.880 0.933 & 1.000 0.999 0.999 & 0.992 0.947 0.969 & 0.994 0.989 0.992 &  0.997 0.940 0.968 \\
    - w/o $LSTM$ & 0.979 0.858 0.915 & 0.993 0.879 0.932 & 1.000 0.999 0.999 & 0.995 0.909 0.951 & 1.000 0.963 0.981 &  0.966 0.953 0.959\\
    - w/o $F_{contx}$ & 0.977 0.060 0.113 & 0.984 0.253 0.403 & 0.999 0.289 0.449 & 0.999 0.242 0.389 & 1.000 0.256 0.407 &  0.842 0.197 0.319 \\
         \bottomrule
    \end{tabular}
    \label{tab:miner-results}
\end{table*}

\begin{table}[t]
\scriptsize
    \centering
    \scriptsize

    \caption{Experiment results for anomaly detection.}
    \begin{subtable}{0.5\textwidth}
    \centering
        \caption{HDFS Dataset.}
        \vspace{-0.1in}
    \label{tab:anomaly-hdfs-results}
    \begin{tabular}{l||c|c|c|c}
    \toprule
    & \multicolumn{4}{c}{Technique}\\
     & DeepLog & LogRobust & CNN & Transformer\\
     Baseline & P \quad R \quad F1 & P \quad R \quad F1 & P \quad R \quad F1 & P \quad R \quad F1\\
    \midrule
    LenMa & .897 .994 .943 & .914  .995  .953 & .924  .995  .958  & .872 .908 .890\\
    AEL & .896  .994  .943 & .935  \textbf{.996}  .964 & .922  .995  .958 & \textbf{.893} .904 .898\\
    Drain & .908  .994  .949 & .934  .994  .963 & .925  .995  .959 & .886 .871 .878\\
    IPLoM & .898  .994  .944 & .940  .994  .966 & .926  \textbf{.996}  .960 &  .889 .904 .896\\
    SemParser & \textbf{.940}  \textbf{.995}  \textbf{.967} & \textbf{.954}  .995  \textbf{.974} & \textbf{.931}  .995  \textbf{.962} & .881  \textbf{.954}  \textbf{.916}\\
        \midrule
    $\Delta$\% & \quad +1.86\% &  \quad +0.82\% &   \quad +0.21\% & +2.00\%\\
    \bottomrule
    \end{tabular}
    \end{subtable}
    
    \begin{subtable}{0.5\textwidth}
    \vspace{0.1in}
    \caption{F-Dataset}
    \vspace{-0.1in}
    \begin{tabular}{l||c|c|c|c}
    \toprule
    & \multicolumn{4}{c}{Technique}\\
     & DeepLog & LogRobust & CNN & Transformer\\
     Baseline & P \quad R \quad F1 & P \quad R \quad F1 & P \quad R \quad F1 & P \quad R \quad F1\\
    \midrule
    LenMa & .717 \textbf{.938} .813 & .714  \textbf{.924}  .806 & .793  .815  .804 & .685  .896  .776 \\
    AEL & .738  .934  .824 & .791  .877  .832 & .747  .924  .826 & .503  \textbf{.962}  .660 \\
    Drain & .824  .867  .845 & .810  .886  .846 & .737  \textbf{.943}  .827 & .693  .919 .790 \\
    IPLoM & .863  .833  .848 & .808  .877  .841 & .834  .834  .834 & .929  .683  .787\\
    SemParser & \textbf{.971}  .927  \textbf{.948} & \textbf{.952}  .913  \textbf{.932} & \textbf{.907}  .899  \textbf{.903} & \textbf{.938}  .904  \textbf{.921}\\
        \midrule
    $\Delta$\% & \quad +11.80\% &  \quad +10.17\% &   \quad +8.27\%&  \quad +16.58\%\\
         \bottomrule
    \end{tabular}

    \label{tab:anomaly-fdataset-results}
    \end{subtable}

\label{tab:anomaly-results}
\end{table}

\subsubsection{Discussion--log parsing comparison}\label{subsec:discussion}
In this section, we discuss why we do not compare~\name~to other syntax-based parsers in the log parsing task where only the templates and parameters are extracted. 
Firstly, the ground-truth for log parsing is not suitable for the semantic parser.
For the logs and their ground-truth templates shown in Table~\ref{tab:discussion-gt} with \Red{highlighted} improper parts,
we observe that ``0'' is not a parameter but a token in the template, because the value for ``CompleteReds'' is always ``0'' in 2000 logs in this template. In contrast, ``0'' will be regarded as an instance in our model, since ``0'' is used to describe ``CompleteReds'' semantically.
Besides, we show how different tokenizer affects the results in the second example, where we consider ``attempt\_14451444'' as an instance for the concept ``TaskAttempt'', but the syntax-based log parsers only regard the number ``14451444'' as parameters, excluding the same prefix ``attempt''. This kind of widely-present distinction occurs 817 times among 2000 logs in the Hadoop log collection.
As a result, it is unfair to compare~\name~with syntax-based parsers in the log parsing task. Instead, we investigate the semantic mining ability in the first research question.

Secondly, log parsing is more of a pre-processing technique for downstream applications rather than an application by itself, and therefore, it will be more meaningful to concern about how the log parsers promote performance in downstream tasks.
For example, if a developer wants to detect anomalies in overwhelming logs, the extracted templates and their parameters are not what he/she needs, but the result from an automated anomaly detection model is. From this perspective, we compare~\name~with four baseline parsers in two log analysis tasks to demonstrate our semantic parser's effectiveness.
On the other hand, our approach could provide
accurate log templates with extra underlying semantics,
so it would naturally promote generalized downstream tasks.

To conclude, \name~is developed as a semantic-based parser instead of syntax-based parser, so the evaluation should be related to its semantic acquisition ability and how the acquired semantics benefits log analysis for downstream tasks in an end-to-end fashion.

\subsection{RQ1: How effective is the~\name~in mining semantics from logs?}\label{subsec:exp-rq1}

In this experiment, we focus on evaluating the \textit{explicit CI pair extraction} in the semantics miner as it serves as a vital step. A high-quality domain knowledge database and further joint parser process could be conducted if and only if the semantics miner extracts high-quality explicit CI pairs from log messages.

Basically, mining the instance-level semantics from log messages is difficult to do with handcrafted rules. Taking logs in Hadoop as examples, there are several ways to describe an instance associated with one concept TaskAttempt:
\begin{itemize}
    \item TaskAttempt: [attempt\_14451444] using containerId ...
    \item attempt\_14451444 TaskAttempt Transitioned from ...
    \item Progress of TaskAttempt attempt\_14451444 is ...
\end{itemize}
The evaluation result across six representative system logs is presented in Table~\ref{tab:miner-results}. Since our work is the first to extract semantics from logs, we do not set baselines for comparison. Other general text mining techniques in the NLP field can only extract keywords (e.g., LDA~\cite{blei2003latent}), but they are not be capable of extracting semantic pairs or parsing log messages to structured templates.
Instead, we conduct ablation studies to explore the effectiveness of each element in the semantics miner, where w/o $F_{char}$, w/o $F_{local}$, w/o $LSTM$ and w/o $F_{contx}$ refers to removing the character-level feature, local word feature, LSTM network, and interval context, respectively. The best F1 score for each system is in bold fonts.

In conclusion, our model could extract not only high quality but also comprehensive instance-level semantics from log messages. We achieve an average F1 score of 0.985 for six systems logs even though we only fine-tune the base model on 50 annotated samples and a large portion of templates are unseen in the test set (the last column in Table~\ref{tab:loghub-stat}).
The promising result indicates our framework has a powerful ability for capturing semantics from log messages.

We attribute the outstanding concept-instance pairs mining ability of~\name~to its comprehensive architectures. The ablation experiments indicate that removing components degrade the performance in varying degrees.
Firstly, to minimize the impact of a large portion of unknown words (e.g., attempt\_14451444) to the model, we devise a character-level feature extraction convolutional network and a local feature extraction method since similar words are always composed of similar character structures. For example, although attempt\_14451444 is different from attempt\_14415371, they share the same structures that the word ``attempt'' following by an underscore and a sequence of numbers.
Secondly, a recurrent network is designed to capture the contextual representation for each word in a sentence, since the same word may have various meanings under different contexts. By removing the bi-LSTM network, words in the sentence are equally regarded as a bag of words.
Thirdly,~\name~naturally learns the patterns between concepts and instances by incorporating the interval context. For instance, if a colon separates two words, the latter word is probably an instance of the prior one, even if the latter one is an unseen word. We find such interval context is quite important, as a dramatic degradation is observed when we remove it.
To conclude, the experiment shows the superiority of the our model by achieving an average F1 score of 0.985 across various system logs.

\subsection{RQ2: How effective is the~\name~in anomaly detection?}\label{subsec:exp-rq2}

To illustrate how~\name~benefits the anomaly detection task, we compare~\name~with four baseline parsers on four different anomaly detection models, and the results are shown in Table~\ref{tab:anomaly-results}. 
Each row represents the performance of four anomaly detection models associated with the selected parser for upstream processing. The last row ($\Delta$) displays how much our semantic parser outperforms the best baseline parser of F1 score, and the negative score indicates how the percentage of ours performs lower than the best baseline. 

In the base HDFS dataset with only 31 templates, although all parsers provides a good performance, we still observe that~\name~also outperform syntax-based parsers by an average F1 score of 1.22\% over four techniques. 
In the more challenging F-Dataset, we observe that~\name~performs at rates approximately above ten percent overall baselines in F1 score, indicating its effectiveness and robustness across various models. 
It outperforms baselines regarding DeepLog, LogRobust, CNN, and Transformer by 11.80\%, 10.17\%, 8.27\%, and 16.58\% respectively, with an average F1 score of 0.926. 
The results on Precision, Recall, and F1 reveal the effectiveness of acquired semantics from logs.

\begin{table*}[t]
\footnotesize
    \centering
        \caption{Experimental results in failure identification task.}
    \begin{tabular}{l||ccc|ccc|ccc|ccc}
    \toprule
    & \multicolumn{12}{c}{Model}\\
    \cmidrule{2-13}
     & \multicolumn{3}{c}{LSTM} & \multicolumn{3}{c}{Atten-biLSTM} & \multicolumn{3}{c}{CNN} & \multicolumn{3}{c}{Transformer}\\
    Baseline & Rec@1 & Rec@2 & Rec@3 & Rec@1 & Rec@2 & Rec@3 & Rec@1 & Rec@2 & Rec@3 & Rec@1 & Rec@2 & Rec@3\\
    \midrule
        LenMa & 0.839 & 0.924 & 0.953 & 0.858 & 0.943 & 0.957 & 0.877 & 0.962 & 0.967 & 0.919 & 0.934 & 0.948 \\
    AEL & 0.844 & 0.919 & 0.953 & 0.853 & 0.915 & 0.962 & 0.810 & 0.905 & 0.929 & 0.858 & 0.929 & 0.953\\
    Drain & 0.844 & 0.919 & \textbf{0.972} & 0.863 & 0.938 & 0.953 & 0.867 & 0.948 & 0.967 & 0.853 & 0.919 & 0.943\\
    IPLoM & 0.848 & 0.943 & 0.957 & 0.863 & 0.948 & 0.962 & 0.867 & \textbf{0.967} & \textbf{0.986} & 0.839 & 0.910 & 0.948\\

    SemParser & \textbf{0.954} & \textbf{0.968} & 0.968 & \textbf{0.954} & \textbf{0.968} & \textbf{0.972} & \textbf{0.945} & 0.963 & 0.972 & \textbf{0.954} & \textbf{0.958} & \textbf{0.968}\\
    \midrule
    $\Delta$\% & +12.50\% & +2.65\% & -0.41\% & +10.54\% & +2.11\% & +1.04\% & +7.75\% & -0.42\% & -1.44\% & +3.81\% & +2.46\% & +2.11\%\\
    \bottomrule
    \end{tabular}
    \label{tab:rca-results}

\end{table*}

\begin{figure}[tb]

    \centering
        {\includegraphics[width=\linewidth]{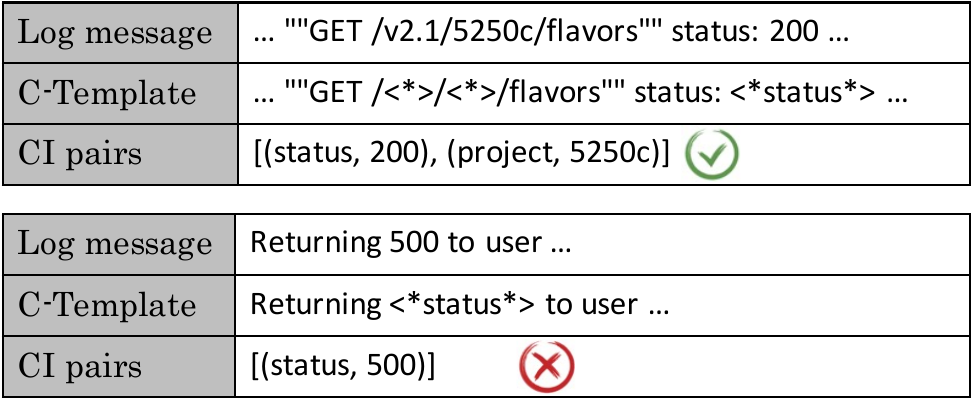}}
             \vspace{-0.1in}
    \caption{A case for anomaly detection.}
    \label{fig:case-anomalydetection}

\end{figure}

We attribute~\name's distinct superiority on its precision to the awareness of semantics we extract, particularly instance-level semantics. Previous studies only use log template sequences to detect anomalies automatically, suffering from missing important semantics.
Taking a case in Figure~\ref{fig:case-anomalydetection} as an example, where C-Template refers to the conceptualized templates. The CI-pairs are either extracted explicitly or implicitly via a domain knowledge database. The green tick indicates a normal log message, while the red cross stands for an anomaly log.
A service maintainer must understand that ``status: 500'' returned by a REST API request reflects the internal server error, while the ``status: 200'' means the request is successful based on ad-hot knowledge. In this way, the maintainer can easily recognize that an API request fails if the return status equals to 500. 
Similarly, feeding semantics like (``status'', ``500'') and (``status'', ``200'') into the anomaly detection model forces the model to learn the relation between ``500'' and ``anomaly'' (or the relation between ``200'' and ``normal''). 
As a result, the model will not mistake a log containing a normal status (e.g., 200) for an anomaly.
The instance-level semantics also resolve problems for unseen logs. Even if the model has never encountered the template before, it is able to correctly predict it as a normal one according to a success status code, and vice versa. Note that without the deliberately established CI Pairs, previous syntax-based parsers cannot distinguish the above normal v.s. anomaly status.

\subsection{RQ3: How effective is the~\name~in failure identification?}\label{subsec:exp-rq3}

\begin{table}[t]
\footnotesize
    \centering
    \footnotesize
 
    \caption{Cases for failure identification.}
 
    \begin{subtable}{0.45\textwidth}
    \centering
    \caption{A case for instance-level semantics.}
        \vspace{-0.1in}
    \begin{tabular}{l||l}
    \toprule
    \textbf{API error} & server add volume\\
            \midrule
    \textbf{Log message} & ... Cannot 'attach\_volume' instance 853cfe1b ... \\
    \textbf{C-Template} & ... Cannot 'attach\_volume' instance $<$*server*$>$ ... \\
    \textbf{CI Pairs} & [(server, 853cfe1b)]\\

         \bottomrule
    \end{tabular}

    \label{tab:rca-case-instance}

    \end{subtable}
    
    \begin{subtable}{0.45\textwidth}
    \centering
    \footnotesize
                \vspace{0.1in}
        \caption{A case for message-level semantics.}
            \vspace{-0.1in}
    \begin{tabular}{l||l}
    \toprule
        \textbf{API error} & network create\\
            \midrule
    \textbf{Log message} &  ... POST /v2.0/networks ... \\
    \textbf{C-Template} & ... POST /$<$*$>$/networks ...\\
    \textbf{Concepts} & [POST, networks]\\
         \bottomrule
    \end{tabular}

    \label{tab:rca-case-message}
    \end{subtable}
    
    
\end{table}

This section demonstrates how effectively our semantic parser enhances failure identification.
The experimental results are shown in Table~\ref{tab:rca-results}, where each row represents the performance with the selected parser and several model architectures. The last row reveals how much~\name~increases the F1 score when compared to the best baseline results. Given that there are 16 types of API errors in F-Dataset, we report Recall@1, Recall@2, Recall@3 score, as we want the top-k suggested errors to cover the real API error.

It is noteworthy that our semantic parser outperforms four baselines by a wide margin, regardless of the analytical techniques. We can observe that our parser surpasses others by 12.5\%, 10\%, 7.75\%, and 3.81\% for LSTM, Atten-biLSTM, CNN, and Transformer in Recall@1, respectively. In general,~\name~shows the promising Recall@1 score of 0.95, indicating the effectiveness of semantics for failure identification.

The impressive performance can be attributed to several reasons.
Firstly, our parser can extract precise conceptualized templates, serving as a basis for downstream task learning. 
We extract conceptualized templates by replacing the instances with their corresponding concepts while reserving all concepts in the template, based on the observation that instances (e.g., time, len, ID) are more likely to be generated in running time. The template number dramatically decreases after conceptualization, giving the sequence of abstract log messages for primitive learning.

Secondly, the instance-level semantics benefits failure identification. 
In the case shown in Table~\ref{tab:rca-case-instance}, ``853cfe1b'' will be regarded as a meaningless character string by the traditional syntax-based parser; however,~\name~recognizes it as a ``server'' from previous log messages. Therefore, the preserved semantics allows the downstream technique to understand that the original log message is talking about the concept \textit{server}, as well as the concept \textit{attach\_volume}, then it will not be hard to infer the API error behind the failure is ``server add volume''.


Thirdly, our parser provides strong messages-level semantics, clues model in resolving failures. For example, Table~\ref{tab:rca-case-message}~shows how the semantic parser extracts the concept ``network'' with the actual API error being ``network create''. With the help of the concept ``network'', the model focuses on network errors and filters other server errors or volume errors. 
To sum up,~\name~benefits the failure identification task by providing message-level semantics and instance-level semantics altogether.



\section{Threat to Validity}\label{sec:threat2validity}

\textbf{Threats to CI pair granularity.}
Our approach can only discover semantic pairs in a single word. For example, for one Zookeeper log ``Connection request from old client /10.10.31.13:40061'', the extracted CI pair is ``(client, /10.10.31.13:40061)'' instead of ``(old client, 10.10.31.13:40061)''. Using ``old client'' is more precise than ``client'' to describe this instance. 
Fortunately, based on our observation, since such multi-word concepts infrequently occur in log messages, using the single-word concept will not alter the semantics too much.

\textbf{Threats to transferability.}
Our model mines semantics relying on manually labeled data. The sampled data for annotation and annotation quality both affect its performance. Fine-tuning with new annotation is required to transfer the model across different systems. In this case, we consider that our model can easily adapt to a new system after fine-tuning with a small amount of data (e.g., Our RQ1 shows that 50 annotated logs are sufficient to transfer a model from OpenStack to Hadoop, with 84.6\% templates in test set are unseen). 

\textbf{Threats to efficiency.} Despite the fact that the neural network used in our approach can effectively mine semantics, it is not as computationally efficient as other statistical parsers. Nevertheless, the issue can be mitigated by batch operation or GPU acceleration. 
Moreover, missing identification of an anomaly can also be very costly. As RQ2 and RQ3 demonstrate \name's effectiveness over other parsers in anomaly detection and failure identification, it is worthy of mining such semantics by sacrificing controllable computational efficiency.
\section{Related Work}\label{sec:relatedwork}

\subsection{Log parsing} 

A series of data mining approaches are proposed for log parsing, which can be further divided into three categories~\cite{he2021survey}: frequent pattern mining, heuristics, and clustering. Among frequent pattern mining approaches, SLCT~\cite{vaarandi2003data} pioneered the automated log parsing, determined whether a token belongs to variables or constants based on its occurrences, assuming that the frequent words are always shown in constants.
Heuristic approaches are more intuitive than others. For example, AEL~\cite{jiang2008abstracting} went over a collection of heuristic rules to conduct log parsing. Another online heuristic log parser Drain~\cite{he2017drain} used a fixed depth parse tree, with each internal leaf node encoding specifically designed parsing rules.
The clustering approaches first encode log messages into vectors, then group the messages with similar vectors. For example, LKE~\cite{fu2009execution} hierarchically clustered messages with a weighted edit distance threshold, then performs group splitting with fine-tuning to extract variables from messages. Another approach LenMa~\cite{shima2016length} encoded each log to its word length vector for clustering.

However, all the above studies only distinguish variables from constants in a log message, assuming the message as a sequence of characters and symbols independent of the variables' meaning. Our work starts from a higher-level semantic perspective, particularly resolves the meaning of parameters and the template in a log message. In this way, our work differs significantly from previous studies. 
One similar Named Entity Recognition (NER) work in log community\cite{shetty2021neural} also noticed the importance of semantics in logs, intending to identify entities in logs. However, the NER task relies on a close-world assumption that all entities are known in advance, suffering the explosion of the number of entity types, which impedes real-world practice and generalization across different systems.
\vspace{-0.05in}
\subsection{Log mining}
Log mining analyzes a large amount of data to facilitate monitoring and troubleshooting software systems~\cite{he2021survey}.
Anomaly detection is a typical log mining task in large-scale software systems, referring to identify logs that do not conform to expected behavior. 
Have encoded the log templates into vectors, previous studies use traditional learning approaches to find anomalies, such as Principal Component Analysis (PCA)~\cite{xu2009largescale}, clustering~\cite{lin2016log}, and Support Vector Machine (SVM)~\cite{liang2007failure}. Some deep learning-based approaches have also been adapted to identify anomalies, such as LSTM~\cite{du2017deeplog, zhang2019robust}, CNN~\cite{lu2018detecting}, Transformers~\cite{nedelkoski2020self} and pre-trained language models~\cite{ott2021robust}. To overcome the unseen log problem, LogRobust~\cite{zhang2019robust} proposed a robust log encoding method with the TF-IDF value and word embeddings, and then an attention-based bi-LSTM was used to learn the importance of each log. 

Although anomaly detection points out whether an anomaly exists in system logs, removing such anomaly requires the help of failure diagnosis. To address the problem, some studies~\cite{jia2017logsed, jia2017approach} automatically constructed time-weighted control flow graphs (TCFG) from normal execution log sequences as the reference model, and then checks the deviation between the reference model and the new coming log sequences to diagnose a failure. 
Finite state models are also used to highlight the difference between the logs~\cite{amar2018using}.
In addition, inspired by the fact that occurred failure manifests recurring, some studies~\cite{jiang2017causes, amar2019mining} were interested in developing a failure matching algorithm to retrieve similar historical failure reports from the report database.
Undoubtedly, as an important part of log analysis and system monitoring, the performances of anomaly detection and fault diagnosis model are affected by the output of upstream tasks. We have demonstrated that our semantic parser enhances the performance of mainstream analysis models.
\section{Conclusion}\label{sec:conclusion}

In this paper, we first point out three limitations of current log parsers: inadequate informative tokens, missing semantics within logs, and missing relation between logs. 
To overcome the limits, we then design~\name, a semantic parser with two phases: a semantics miner aiming to mine explicit semantics from logs, as well as a joint parser leveraging domain knowledge to infer implicit semantics.
We then conduct extensive experiments to evaluate~\name~in six representative system logs for semantic mining ability, which achieves an average F1 score of 0.985. Moreover, we evaluate our approach in two downstream log analysis tasks (i.e., anomaly detection and failure identification). The experimental results demonstrate that our method outperforms syntax-based log parsers by large margins, confirming the importance of understanding semantics in log analysis. We release code and data for future research\footnote{Please find in https://github.com/YintongHuo/SemParser.}.
\section*{ACKNOWLEDGEMENT}
The work described in this paper was supported by the National Natural Science Foundation of China (No. 62202511), and the Research Grants Council of the Hong Kong Special Administrative Region, China (No. CUHK 14206921 of the General Research Fund).
\newpage
\balance
\bibliographystyle{IEEEtran}
\bibliography{semparser}

\end{document}